\documentclass[journal,twocolumn]{IEEEtran}

\title{Design of Non-Coherent and Coherent Receivers for Chirp Spread Spectrum Systems
\thanks{}
\thanks{}
}
\author{Tung T. Nguyen and Ha H. Nguyen, \emph{Senior Member, IEEE}
\thanks{The authors are with the Department of Electrical and Computer Engineering, University of Saskatchewan, 57 Campus Drive, Saskatoon, SK, Canada S7J 4E3. Emails: {\tt tung.nguyen@usask.ca, ha.nguyen@usask.ca}.}
\thanks{This work supported by NSERC/Cisco Industrial Research Chair in Low-Power Wireless Access for Sensor Networks.}}

\usepackage{amssymb,amsmath}
\usepackage[final]{graphicx}
\usepackage{cite,bm}
\usepackage{balance}
\usepackage[eulergreek]{sansmath}
\usepackage{enumerate}
\usepackage{setspace}
\usepackage{color}
\usepackage{times}
\usepackage{caption}
\usepackage{multirow}
\usepackage{subcaption}

\newcommand{\nn}{{\nonumber}}

\newcommand{\comment}[1]{}

\begin{document}

\maketitle

\begin{abstract}
LoRaWAN is a prominent communication standard to enable reliable low-power, long-range communications for the Internet of Things (IoT). The modulation technique used in LoRaWAN, commonly known as LoRa modulation, is based on the principle of chirp spread spectrum (CSS). While extensive research has been conducted on improving various aspects of LoRa transmitter, the design of LoRa receivers that can operate under practical conditions of timing and frequency offsets is missing. To fill this gap, this paper develops and presents detailed designs of timing, frequency and phase synchronization circuits for both non-coherent and coherent detection of CSS
signals. More importantly, the proposed receiver can be used to detect the recently proposed CSS-based modulation that embeds extra information bits in the starting phases of conventional CSS symbols. Such a transmission scheme, referred to as phase-shift keying CSS (PSK-CSS) helps to improve the transmission rates of the conventional CSS system. In particular, it is shown that the bit error rate performance of the PSK-CSS scheme achieved with the proposed practical coherent receiver has only 0.25 dB gaps as compared to the ideal co-coherent receiver.
\end{abstract}

\section{Introduction}\label{intro}

In the last ten years or so, Internet-of-Things (IoT) has become a main stream in communications technology. By connecting billions of devices to the Internet, IoT allows collecting and sharing a massive amount of data with ease. IoT enables objects to perform various tasks without human interaction. A ``thing'' in an IoT network can be a farm animal carrying a biochip transponder, a home security sensor that alerts its owner when there is a break-in, a heart monitor implanted inside a person, or anything else that can be assigned an address and is able to transfer data over Internet \cite{Fuq15,Zan14}.

In many cases, the ubiquitous application of IoT requires exceptional energy efficiency of the end nodes (end devices), where battery lifetime is expected to last for many years. Moreover, business opportunities for new
services in various fields such as logistic tracking, traffic control, environmental monitoring and personal healthcare \cite{Gha19} demand reliable communications over a long range (in the order of ten kilometers). These low power and long range requirements have led to the development of Low-Power Wide-Area Networks (LPWANs) solutions for IoT applications. Different from short-range wireless networks such as Bluetooth, Zigbee and Z-Wave (which are designed to cover very short distances with minimal power
consumption) and cellular networks (which are geared towards very high data rate transmission and with higher transmission power), LPWANs trade transmission rate to achieve very large coverage and with excellent energy efficiency \cite{Raz17}. Prominent LPWAN technologies include Ingenu \cite{Ingenu}, DASH \cite{Weyn15}, LoRaWAN \cite{lorawan,Weyn15} and SigFox \cite{Sigfox}. These LPWANs are expected to co-exist in order to provide connectivity for billions of IoT devices \cite{Centenaro16,Augustin16}.

This paper is specifically concerned with the physical layer of LoRaWAN. This technology is gaining tremendous commercial growth in more than 100 countries around the world\footnote{https://lora-alliance.org/}. The physical layer of LoRaWAN is built on the type of modulation known as chirp spread spectrum (CSS), which is also commonly referred to as LoRa modulation \cite{Rey16,Vangelista17}. In this paper, the terms CSS and LoRa are used interchangeably. CSS is known for its great flexibility in providing a trade-off between reception sensitivity and data rate. Spreading factor
(SF) is the most important parameter in CSS modulation. Increasing SF can
significantly extend the communication range, but it comes at the cost of
a lower transmission rate. Bandwidth is another adjustable parameter in LoRaWAN. As expected, using a larger bandwidth enhances the transmission rate and, at the same time, provides better immunity to narrow-band noise and ingress. Evaluation of link performance, as well as system-level performance of LoRaWAN can be found in \cite{Feltrin18,Nguyen19}.

Since the introduction of LoRaWAN, there has been active research in improving various aspects of LoRa modulation. For example, an efficient design of the LoRa transmitter is presented in \cite{Nguyen19,TungUS21} that eliminates the use of ROM and also introduces pulse shaping to improve the
spectral compactness of LoRa signals. Since having low data rates is considered a major disadvantage of the conventional LoRa modulation, a large research effort was also devoted to improving the data rates of LoRaWAN.

Using the starting phases of CSS symbols to carry additional information bits is explored in \cite{Nguyen19,Bom19} and the resulting scheme is called phase-shift keying CSS (PSK-CSS), or phase-shift keying LoRa (PSK-LoRa). In \cite{Almeida20}, superposition of in-phase and quadrature CSS signals is introduced to double the spectral efficiency of the conventional LoRa. The authors in \cite{InterleavedCSS} propose to use interleaved chirps along with linear up chirps to double the number of chirp signals, hence increasing the data rate by one bit per each symbol. Another approach
to double the number of chirp signals is introduced in \cite{SSKLoRa} in which down chirps are used instead of the interleaved chirps. It is demonstrated in \cite{SSKLoRa} via correlation and bit error rate (BER) analysis that using down chirps instead of interleaved chirps reduces the peak correlation between the added signal set and the original signal set, hence improving the BER performance. More recently, the concept of index modulation has been applied
to chirps in \cite{HanifUS20,hanif2021frequencyshift}, which leads to a novel CSS-based modulation scheme that can provide much higher data rates than the conventional LoRa modulation scheme.

While much research has been conducted on improving the conventional LoRa
modulation and transmitter, detailed studies and design of \emph{practical} LoRa receivers are missing in the literature. The commercialized LoRa physical-layer solution is currently patented by Semtech \cite{SemtechAN22,Horn10} and not described in detail \cite{sforza13}. In all the research papers discussed above, the LoRa receivers operate under the idealistic
assumption of having \emph{perfect} synchronization, i.e., having no timing and frequency errors at the receiver \cite{sforza13,Rey16,Ouyang17,Van17,Mar19}. In practice, however, the timing error is unavoidable due to the imperfection in the preamble detection process, whereas the frequency error happens due to the fact that crystal oscillators at the transmitter
and the receiver are physically separated and different. In reality, synchronization plays a crucial role in the final detection of the information bits. Large synchronization errors, if not accounted for, leads to severe interference (non-orthogonality) among chirps, and consequently unacceptable/unreliable detection performance.

Against the above background, this paper develops and presents a detailed
design of  practical non-coherent and coherent receivers for CSS modulation. In particular, the proposed design includes circuits for timing, frequency and phase synchronization. The coarse timing and frequency synchronization is accomplished based on the preamble and can be used in a non-coherent receiver. Moreover, fine timing, frequency and phase synchronization is achieved using data symbols in the CSS burst. It is emphasized that, with fine synchronization, the proposed design not only enables coherent detection of conventional CSS modulation under practical scenarios of having timing and frequency offsets, it allows a practical implementation of the PSK-CSS scheme proposed in \cite{Nguyen19,Bom19} to provide higher
data rates than that of the conventional CSS modulation scheme. Specifically, once the receiver's phase is locked, the PSK modulated bits can be recovered by examining the phase of the discrete-time Fourier transform (DFT) bin having the highest magnitude. The main ideas and algorithms presented in this paper were originally described in \cite{CSS-TxRx}.

The remainder of this paper is organized as follows. Section \ref{sec-TX}
reviews the transmitter of the conventional CSS system, as well as both non-coherent and coherent detection methods under the ideal channel condition, i.e., without any timing and frequency offsets between the transmitter and the receiver. It also reviews the PSK-CSS scheme proposed in \cite{Nguyen19,Bom19} to increase the data rates of the conventional CSS system. Section \ref{sec-RX} presents the proposed design of practical non-coherent and coherent receivers, which includes circuits for timing, frequency and phase synchronization. Simulation results and discussion are provided in Section \ref{sec-res}. Section \ref{sec-con} concludes the paper.


\section{Conventional CSS and PSK-CSS Systems}\label{sec-TX}

\subsection{Conventional CSS Transmitter}

CSS or LoRa modulation is built on a set of $M$ chirps, which are signals
whose frequency linearly increases or decreases over time. All $M$ chirps
are orthogonal over the symbol duration $T_{\rm sym}$ and distinguished by properly chosen starting frequencies. This means that each chirp, or LoRa symbol, can carry ${\rm SF}=\log_2 M$ bits, which are also known as \emph{spreading factor}. In practice, ${\rm SF}$ could take values in the
set $\{7, 8, \ldots, 12\}$.

Let $B$ be the bandwidth and $T_s = 1/B$ the sampling period. Then each
LoRa symbol can be represented by exactly $M = T_{\rm sym}/T_s = 2^{\textup{SF}}$ samples. Furthermore,
the baseband discrete-time basic up chirp (of length $M$ samples) is given as \cite{Nguyen19}:
\begin{equation}\label{x0n}
x_0[n]=\exp\left(j2\pi\left(\frac{ n^2}{2M}-\frac{n}{2}\right)\right),\; n=0,1\dots, M-1.
\end{equation}
Then, the set of $M$ orthogonal chirps can be simply constructed from $x_0[n]$ as $x_m[n]=x_0[n+m]$, $m=0,1\dots, M$.

\begin{figure*}[htb!]
\centering
\includegraphics[scale=0.80]{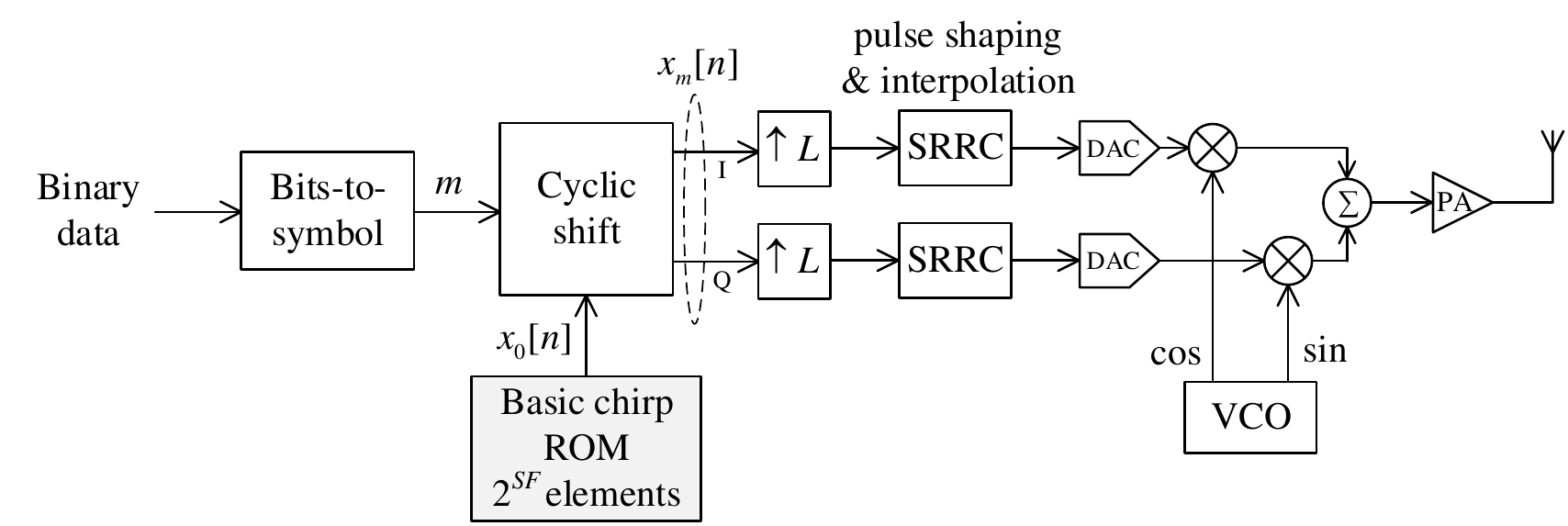}
\caption{Block diagram of a LUT-based LoRa transmitter with spectral shaping.}
\label{fig-css-tx}
\end{figure*}

Fig.~\ref{fig-css-tx} shows the block diagram of a conventional lookup-table (LUT) based LoRa transmitter. The most important component of the transmitter is a ROM that stores $M$ samples of the basic chirp in \eqref{x0n}. The binary data is converted to a stream of LoRa symbols, each carrying ${\rm SF}$ bits. Mapping from a LoRa symbol to a transmitted chirp is performed by cyclically shifting the basic chirp by an amount equivalent to the symbol value $m$, producing a corresponding digital complex baseband chirp $x_m[n]$. Each digital baseband chirp is then up-sampled (by a factor of $L$), spectrally shaped by a pair of square-root raised cosine (SRRC) filters before being converted to an analog baseband signal. The benefit of implementing SRRC filters in LoRa transmitter is demonstrated in
\cite{Nguyen19}. Furthermore, a very efficient implementation of the LoRa
transmitter that does not require a ROM is also presented in \cite{Nguyen19}. Since the focus of this paper is on the receiver design, the LUT-based transmitter is used for ease of explanation.

\subsection{Detection of CSS Signals under Perfect Synchronization}

Before presenting the proposed design of non-coherent and coherent receivers for LoRa signals in the next section, it is relevant to review the detection algorithms of LoRa signals under \emph{perfect} timing and frequency synchronization that are widely adopted in the literature.

Without the use of SRRC filters and under perfect timing and frequency synchronization and additive white Gaussian noise (AWGN), the baseband received signal (after being down-converted and sampled) can be expressed as
\begin{equation}
y_m[n] = \exp(j \psi)x_m[n] + w[n], \; n = 0,1,\ldots,M-1.
\end{equation}
In the above expression, $\psi$ is a random phase rotation caused by the fact that the oscillators at the transmitter and receiver are not phase-locked, and $w[n]$ represents  AWGN samples that are independent and identically-distributed with zero mean and variance $N_0$. For such an input/output model, it is simple to see that the signal-to-noise ratio (SNR) is $\text{SNR}~=~\frac{1}{N_0}$.

Detection of LoRa signals starts with a multiplication with the complex conjugate of the basic up chirp, a process known as \emph{de-chirping}. This yields
\begin{eqnarray} \label{vmn}
u_m[n]&=& y_m[n]  x^*_0[n] \nn \\
&=& \underbrace{\exp\left\{j \psi_m\right\}}_{\text{constant phase}} \underbrace{\exp \left\{ \frac{ j 2 \pi  nm}{M}  \right\}}_\text{linear phase} + \hat{w}[n],\nn\\  &&\qquad n=0,1,\ldots, M-1,
\end{eqnarray}
where the constant phase term is $\psi_m = \psi + 2 \pi \left( \frac{m^2}{2M} - \frac{m}{2}\right)$ and $\hat{w}[n]=w[n] x^*_0[n]$ is also a Gaussian noise sample with zero mean and variance $N_0$. Thus, in the absence of noise, the received signal after de-chirping is a pure sinusoid with a frequency of ${m}/{M}$ (cycles/sample). This suggests performing $M$-point DFT on the de-chirped signal to obtain
\begin{eqnarray}
\lefteqn{v_m[k] = \frac{1}{\sqrt{M}}\sum_{n=0}^{M-1} u_m[n] \exp\left\{ \frac{-j 2 \pi n k}{M}  \right\}}\nn\\ 
& & = \frac{1}{\sqrt{M}}\sum_{n=0}^{M-1} \exp\left\{j \psi_m \right\}
\exp \left\{ \frac{ j 2 \pi  nm}{M}  \right\} \exp \left\{ \frac{-j 2 \pi
n k}{M} \right\} \nn\\
&& +
\underbrace{\frac{1}{\sqrt{M}} \sum_{n=0}^{M-1} \hat{w}[n] \exp \left\{
\frac{-j 2 \pi n k}{M} \right\}}_{W[k]} \nn \\
& & = \frac{\exp\left\{j \psi_m \right\}}{\sqrt{M}}  \sum_{n=0}^{M-1}
\exp \left\{ \frac{ j 2 \pi  n(m-k)}{M}  \right\} + W[k]    \nn \\
& & = \begin{cases} \sqrt{M} \exp\left\{j \psi_m \right\} + W[m], \quad
\text{if } k = m \\W[k], \quad \text{otherwise}
\end{cases}
\end{eqnarray}
The above analysis shows that all the power in the de-chirped signal is concentrated at a single frequency bin, namely the $m$th bin if the transmitted LoRa symbol is $m$, whereas all the other $M-1$ bins contain only noise. At this point, it is useful to define a peak SNR (PSNR) as
\begin{equation} \label{psnr}
\text{PSNR} = \frac{ \mathbb{E} \left\{ \left\vert \sqrt{M} \exp\left\{j \psi_m \right\}\right\vert^2 \right\} }{\vert W[k] \vert^2}=\frac{M}{N_0},
\end{equation}
which indicates the difference in terms of power between the ``power'' bin versus the rest (``noise'' bins). It is obvious that a higher PSNR would lead to a better chance of detecting the transmitted symbol correctly. The ratio between PSNR and SNR, which is $M$ in linear scale, is called the \emph{processing gain}. Using a higher SF would lead to a higher processing gain and thus improve the performance of the transmission, but at the cost of a lower transmission rate.

Finally, the transmitted LoRa symbol can be detected coherently or non-coherently. If the receiver can be phase-locked to the transmitter (which requires a phase-locked loop (PLL) at the receiver), the phase offset $\psi$ can be estimated and the constant phase term $\psi_m$ can be computed for each LoRa symbol $m$. It follows that the coherent detection can be performed as $\hat{m}=\underset{k=0,\ldots,M-1}{\text{argmax}} v_m[k] \exp\{ -j \psi_m \}$. On the other hand, non-coherent detection does not require to estimate $\psi$ and is performed as $\hat{m}=\underset{k=0,\ldots,M-1}{\text{argmax}}\left\vert v_m[k] \right\vert$.

\subsection{PSK-CSS Scheme}

Performance analysis and comparison of both coherent and non-coherent detection algorithms under perfect timing, frequency and phase synchronization are presented in \cite{Nguyen19} for the conventional CSS system. While performing slighter better than con-coherent detection, the extra complexity required by coherent detection does not justify its use in existing
LoRa systems. However, the situation is different when there is a need to
increase the data rates of existing LoRa systems by means of embedding additional information bits into the starting phases of LoRa symbols. This scheme is known as PSK-CSS or PSK-LoRa \cite{Nguyen19,Bom19}.

Specifically, using a $Q$-ary PSK constellation, a transmitted chirp (symbol) in the PSK-CSS system is defined using two values: the symbol number
$m$, $0\leq m\leq M-1$ (as in the conventional CSS), and the phase symbol
$p$, $0\leq p\leq Q-1$. The discrete-time baseband samples of such a transmitted chirp are simply $x_m[n] \mathrm{exp}\left\{j 2\pi (p/Q)\right\} =  x_0[n+m] \mathrm{exp}\left\{j 2\pi (p/Q)\right\}$. It is obvious that for such an improved design of the PSK-CSS system, the use of a coherent receiver is a must. An efficient design of a practical coherent receiver, including timing, frequency and phase synchronization, is presented in
the next section.

\section{Proposed Design of Non-Coherent and Coherent Receivers}\label{sec-RX}

Any practical receiver for digital communications over radio-frequency (RF) needs to have a timing recovery circuit. Furthermore, a coherent receiver also requires phase and frequency recovery circuits. In this section,
we first present a method to detect the timing error by processing the received preamble symbols under perfect frequency synchronization (Section \ref{sec-timing}). The method is then extended in Section \ref{sec-timing-freq} to detect both timing and frequency errors, and hence serves as coarse timing and frequency synchronization that is based on preamble symbols only. The developed coarse timing and frequency synchronization circuits are employed in the proposed non-coherent receiver. Then Section \ref{sec-fine} presents fine timing, frequency and phase synchronization that is based on data-directed symbols. Such fine synchronization is employed in the proposed coherent receiver to achieve excellent BER performance of
the PSK-CSS scheme.

\subsection{Timing Estimation Under Perfect Frequency Synchronization}\label{sec-timing}

A data burst sent from the transmitter to the receiver always starts with
a preamble, which consists of one or multiple basic down chirps, followed
by data symbols. To aid timing detection, the basic down chirp is used to
create a preamble sequence. A basic down chirp is simply the complex conjugate of the basic up chirp,
given as
\begin{equation}
x_0^*[n] = \exp \left\{-j 2 \pi \left( \frac{n^2}{2M} - \frac{n}{2} \right)\right\}, \; n = 0,1, \ldots, M-1
\end{equation}

\begin{figure*}[htb!]
\centering
\includegraphics[width=0.850\textwidth]{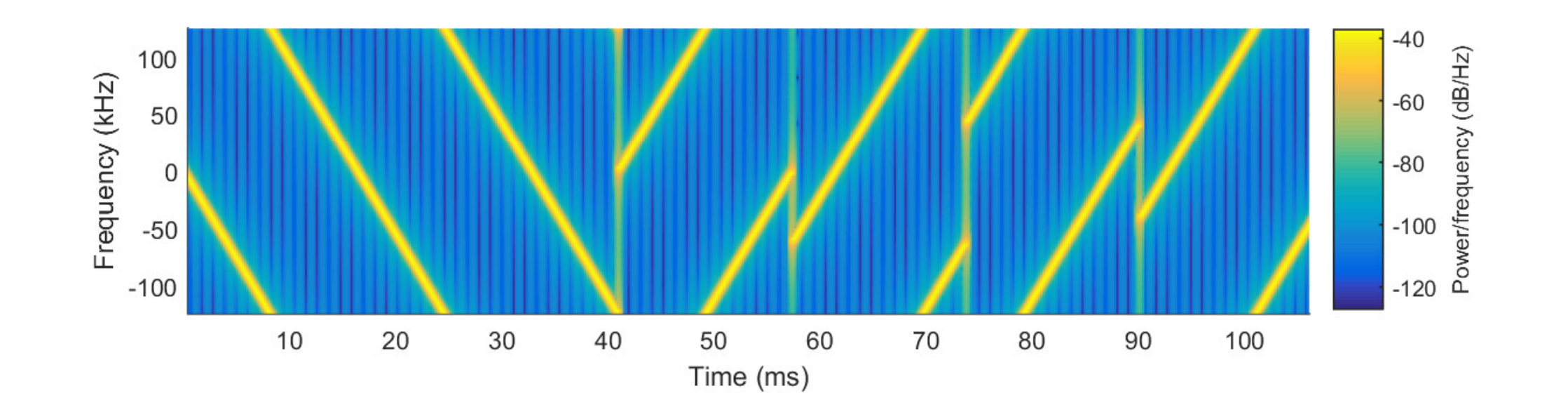}
\caption{Example of a preamble with $N_{\rm CP} = M/2$ and $N_{\rm PR} = 2$.}
\label{fig-preamble}
\end{figure*}

The processing of LoRa signals is similar to that of OFDM signals in the sense that the received signal is processed block-by-block (i.e., symbol-by-symbol) where each block (i.e., symbol) consists of $M$ samples. Typically at the receiver, there is a burst detector that detects the beginning of a LoRa burst based on a sudden rise in the incoming signal power. In
other words, the burst detector is a \emph{coarse} timing detector. Because the coarse timing detector is not so accurate, there would be inter-symbol-interference (ISI) in the obtained preamble samples, i.e., each obtained preamble might be affected by data samples. However, the ISI can be avoided by appending a cyclic prefix on the preamble symbol, or by simply
repeating the preamble symbol multiple times. Repetition guarantees that only the first received preamble symbol might suffer ISI but the subsequent preamble symbols do not. The repetition also gives better preamble detection under the influence of noise. The preamble sequence can be written
as
\begin{multline}
p[n] = x_0^*[(n-N_{\rm CP}) \mod M], \\
  n = 0, 1, \ldots, MN_{\rm PR}+N_{\rm CP}-1,
\end{multline}
where $N_{\rm CP}$ is the number of cyclic prefix samples and $N_{\rm PR}$ is the number of repeated basic chirps (symbols) in the preamble sequence. Fig.~\ref{fig-preamble} shows an example of a LoRa preamble with $N_{\rm CP} = M/2$ and $N_{\rm PR} = 2$, followed by 4 data symbols.

With the implementation of pulse shaping filters, the received preamble after removing CP can be shown to be
\begin{multline} \label{rn0}
r[n] = \sum_{l=0}^{L-1} \hat{h}[l] p[n + N_{\rm CP}-l] + \eta[n], \\
	n = 0, 1, \ldots, MN_{\rm PR}+N_{\rm CP}-1,
\end{multline}
where $\eta[n]$ is an AWGN sample and $\hat{h}[l]$ represents the \emph{composite} channel consisting of all filters between the transmitter and the receiver. Specifically,
\begin{equation}
\hat{h}[l] = h_{\rm tx}[l] * h_{\rm up}[l] * h_{\rm ch}[l] * h_{\rm down}[l] * h_{\rm rx}[l]
\end{equation}
where $h_{\rm tx}[l]$ is the pulse shaping filter at the transmitter (which is an SRRC filter), $h_{\rm up}[l]$ is the interpolation filter at the
transmitter (to remove aliases caused by upsampling), $h_{\rm ch}[l]$ is
the discrete-time impulse response of the channel, $h_{\rm down}[l]$ is the decimation filter at the receiver (to reject alias prior to downsampling), and $h_{\rm rx}[l]$ is the matched filter at the receiver (which is
also an SRRC filter). Because LoRa signals have narrow bandwidths (125, 250, or 500 kHz), the channel is assumed to have a constant frequency response and can be represented as a pure delay, $h_{\rm ch}[l] = \delta(l - \tau)$, where $\tau$ represents the composite effect of the channel delay and coarse timing estimation error. Note that $\tau$ has both integer and fractional components.

Since the pair of SRRC filters have the narrowest bandwidth in the system
and because the combination of the decimation filter, $h_{\rm down}[l]$, and the interpolation filter, $h_{\rm up}[l]$, would yield a flat spectrum over the bandwidth of the SRRC filters, the composite channel is well approximated as
\begin{equation}
\hat{h}[l] \approx h_{\rm RC}(l - \tau),
\end{equation}
where $h_{\rm RC}(l)$ is the impulse response of a raised cosine (RC) filter, resulted by the convolution of the two SRRC filters, $h_{\rm tx}[l]$
and $h_{\rm rx}[l]$. The impulse response of the RC filter is
given as
\begin{multline}
h_{\rm RC}(l) = \frac{\sin(\pi l ) }{ \pi l} \frac{\cos(\beta \pi l)}{ 1 - (2 \beta l)^2  }, \; l = 0, 1, \ldots, L-1
\end{multline}
where $\beta$ is the roll-off factor and $L$ is the truncated length of the RC filter. Substituting the above approximation of $\hat{h}[l]$ into \eqref{rn0} has
\begin{multline} \label{rn}
r[n] = \sum_{l=0}^{L-1} h_{\rm RC}(l - \tau) p [n + N_{\rm CP}-l] + \eta[n] \\
n = 0, 1, \ldots, MN_{\rm PR}+N_{\rm CP}-1,
\end{multline}

Assuming $N_{\rm CP} \geq L$, i.e., \emph{sufficient} cyclic prefix, there is no ISI in  preamble calculation. Moreover, to reduce the effect of noise, the $N_{\rm PR}$ consecutive preamble symbols in the preamble sequence are accumulated into one block of $M$ samples for preamble processing, i.e.,
\begin{eqnarray} \small \label{yn}
& y[n] &= \frac{1}{N_{\rm PR}} \sum_{b=0}^{N_{\rm PR}-1} r[n + bM] ,\quad n=0,1,\ldots, M-1 \nn \\
& &\hspace{-1cm} = \sum_{l=0}^{L-1} h_{\rm RC}(l - \tau) x_0^*[n-l] +
\frac{\sum_{b=0}^{N_{\rm PR}-1}\eta[n+bM]}{N_{\rm PR}}. \label{snrpr}
\end{eqnarray}
Because the noise samples are uncorrelated, the noise power is reduced by
a factor of $N_{\rm PR}$ by averaging. As a result, the SNR in preamble processing is proportional to the number of repeated preamble symbols used
in the preamble sequence.

The accumulated preamble symbol is then de-chirped to obtain
\begin{eqnarray}  \label{uyx}
& u[n] &= y[n] x_0[n] = \sum_{l=0}^{L-1} h_{\rm RC}(l - \tau) x_0^*[n-l] x_0[n]  + \hat{\eta}[n]  \nn \\
& &\hspace{-1cm}= \sum_{l=0}^{L-1} h_{\rm RC}(l - \tau) \exp \left\{j
2 \pi \left( -\frac{(n-l)^2}{2M} + \frac{n-l}{2} \right) \right\} \nn \\
& & \hspace{2cm}  \times \exp \left\{j 2 \pi \left( \frac{n^2}{2M} - \frac{n}{2} \right) \right\} + \hat{\eta}[n] \nn \\
& &\hspace{-1cm}=  \sum_{l=0}^{L-1} h_{\rm RC}(l - \tau) \exp\left\{-j \pi \left(\frac{l^2}{M} + l \right) \right\} \nn \\
& & \hspace{3cm}  \times \exp\left\{\frac{j 2 \pi nl}{M} \right\} + \hat{\eta}[n].
\end{eqnarray}

The de-chirping process is followed by an $M$-point DFT, whose output is
\begin{eqnarray} \label{vk}
&v[k] &= \frac{1}{M}\sum_{n=0}^{M-1} u[n] \exp\left\{ \frac{-j 2 \pi nk}{M} \right\} \nn \\
& &= \frac{1}{M}\sum_{n=0}^{M-1} \sum_{l=0}^{L-1} h_{\rm RC}(l - \tau) \exp\left\{-j \pi \left(\frac{l^2}{M} + l \right) \right\} \nn \\
& & \hspace{3cm} \times \exp\left\{\frac{j 2 \pi n(l-k)}{M} \right\} + \bar{\eta}[k] \nn \\
& &= \frac{1}{M}\sum_{l=0}^{L-1} h_{\rm RC}(l - \tau) \exp\left\{-j \pi \left(\frac{l^2}{M} + l \right) \right\}  \nn \\
& & \hspace{2cm} \times \sum_{n=0}^{M-1} \exp\left\{\frac{-j 2 \pi n(k-l)}{M} \right\}+  \bar{\eta}[k]  \nn \\
& &= h_{\rm RC}(k - \tau) \exp\left\{-j \pi \left(\frac{k^2}{M} + k \right) \right\} + \bar{\eta}[k].
\end{eqnarray}

A very important and useful observation of \eqref{vk} is that, in the absence of noise, the magnitude of the DFT samples follows the magnitude of the RC pulse, i.e., $\big\vert v[k] \big\vert \approx\big\vert h_{\rm RC}(k - \tau) \big\vert$. This suggests that the integer timing delay can be
detected by simply locating the peak magnitude of $v[k]$:
\begin{equation} \label{tau_int}
\tau_{\rm int} = \underset{k}{\text{argmax}} \vert v[k] \vert.
\end{equation}

Next, the fractional timing delay can be found by obtaining two additional samples at half-a-sample delay on both sides of the peak, denoted as $v[\tau_{\rm int}-0.5]$ and $v[\tau_{\rm int}+0.5]$. How these samples can be obtained is explained in detail at the end of this section.  Using these three samples, parabolic interpolation can be performed to estimate the true location of the peak. Specifically, the fractional timing delay is
obtained as:
\begin{equation} \label{tau_frac}
\tau_{\rm frac} = \frac{ \vert v[\tau_{\rm int}-0.5] \vert - \vert v[\tau_{\rm int}+0.5]\vert}{ 4 (\vert v[\tau_{\rm int}-0.5] \vert + \vert v[\tau_{\rm int}+0.5]\vert - 2 \vert v[\tau_{\rm int}]\vert ) }
\end{equation}

Finally combining \eqref{tau_int} and \eqref{tau_frac} yields the absolute timing delay as
\begin{equation}
\tau = \tau_{\rm int} + \tau_{\rm frac} \quad ({\rm UI}),
\end{equation}
where UI stands for unit-interval of timing delay, which is equivalent to
the sampling period $T_s=1/B$ in this paper.

\begin{figure}[thb!]
\centering
\includegraphics[width=0.45\textwidth]{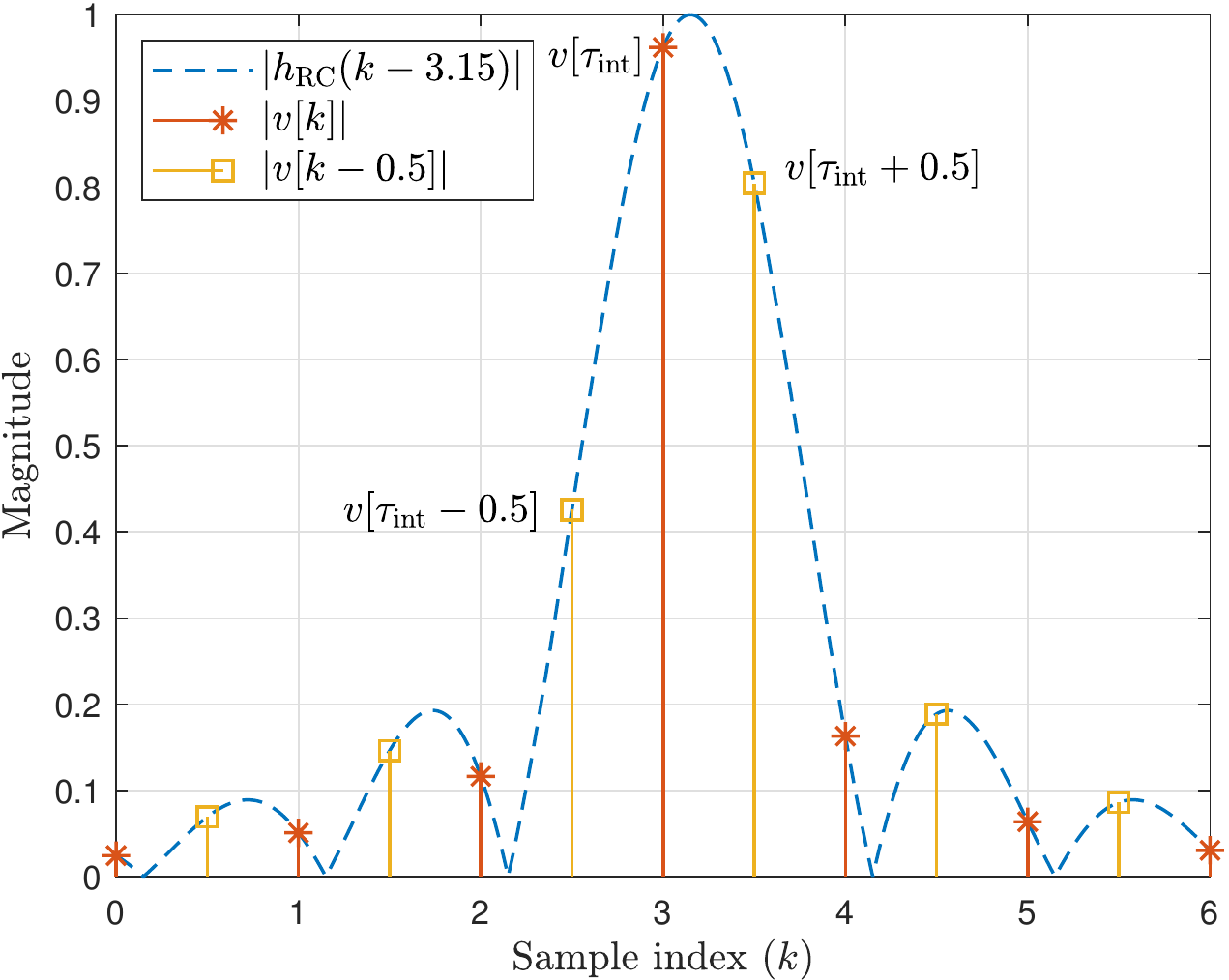}
\caption{Timing detection using 3-point interpolation.}
\label{fig-rc_pulse}
\end{figure}

Fig. \ref{fig-rc_pulse} illustrates the idea of using 3 samples around the peak magnitude to detect the true peak of an RC pulse, and hence the timing error. It can be seen that, if the fractional delay is not accounted
for, the signal power spreads out to multiple taps causing SNR reduction.
Therefore it is very important to compensate for any timing error so that
all the signal power is concentrated in a single tap, namely $v[\tau_{\rm
int}]$, whose location determines the demodulated symbol.

\begin{figure*}[htb!]
\centering
\includegraphics[scale=0.70]{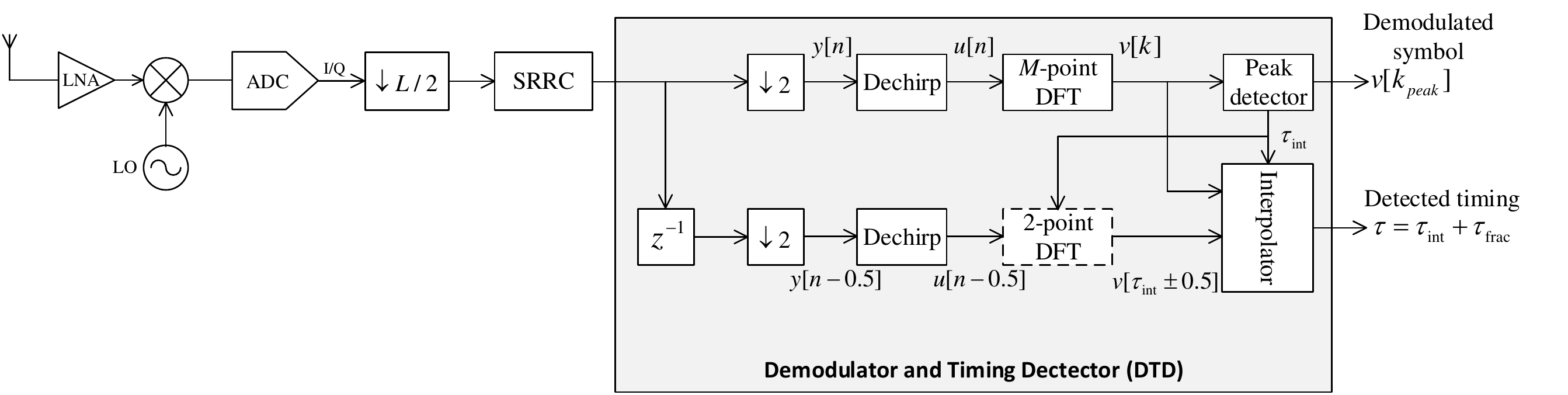}
\caption{Block diagram of the demodulator and timing detector under ideal
frequency synchronization.}
\label{fig-detector}
\end{figure*}

Fig. \ref{fig-detector} shows a detailed block diagram of a CSS demodulator with the proposed timing detector (under ideal frequency synchronization). The received RF signal is first amplified using a low-noise amplifier (LNA), then mixed with a local oscillator (LO) to move the analog signal to baseband. The baseband analog signal is converted to digital using an analog-to-digital converter (ADC). Corresponding to the implementation at the transmitter, the digital signal at the output of the ADC has an up-sampling factor of $L$. This digital signal is then down-sampled to two times the baseband rate before processed by the SRRC filter to limit the noise bandwidth and reject ISI.

After the SRRC filter, the signal is split into two branches: the first branch performs CSS symbol demodulation and the second branch detects the timing of the received symbol. As shown in the figure, the symbol timing detection uses some of the information obtained from the first branch. The two branches are included inside a block named ``demodulator and timing
detector'' (DTD). In the first branch, the signal is further down-sampled
to baseband to obtain $y[n]$. The signal $y[n]$ is then de-chirped and processed with the $M$-point DFT, as expressed in \eqref{yn}, \eqref{uyx} and \eqref{vk}, respectively. A peak detection is performed on $v[k]$ to demodulate the symbol and to obtain the integer delay $\tau_{\rm int}$ as
in \eqref{tau_int}. Compared to the signal in the first branch, the signal in the second branch is offset by one sample delay. Since the single sample delay is performed at twice the baseband rate, after down-conversion
by a factor of two, the baseband sequence effectively experiences half-a-sample delay. With a slight abuse of notation, we denote $y[n-0.5]$ as a sequence having half-a-sample delay with respect to $y[n]$, which is given as
	\begin{equation} \small \label{yn05}
	y[n-0.5] = \sum_{l=0}^{L-1} h_{\rm RC}(l - \tau - 0.5) x_0^*[n-l] + \text{noise}.
	\end{equation}
Performing the same signal processing operations as in \eqref{yn}--\eqref{vk} on $y[n\pm 0.5]$, we obtain $v[k\pm 0.5]$ after the DFT.

It is pointed out that, since only two samples around the peak, i.e., $v[\tau_{\rm int} \pm 0.5]$, are needed, the second branch does not need a full-size $M$-point DFT, but rather a 2-output $M$-point DFT only. This should reduce the computational complexity of the timing detector.

\subsection{Joint Timing and Frequency Estimation for the Proposed Non-Coherent Receiver}\label{sec-timing-freq}

Besides the timing error, there is always a frequency error (or frequency
offset) between the transmitter and the receiver. The amount of error depends on the accuracy of the local oscillator (LO), which is specified in part-per-million, or ppm. The typical accuracy of an LO is about $\pm 20$
ppm, which is equivalent to having a frequency offset of up to 18.3 kHz between the transmitter and the receiver if the LO frequency is set at 915
MHz. The frequency offset, normalized to the signal bandwidth $B$ and the
spreading factor, is defined as
\begin{equation}
\epsilon = \frac{M \Delta_f}{B} \quad ({\rm UI}),
\end{equation}
where $\Delta_f$ is the offset in Hz. For example, considering a system with ${\rm SF}=10$ and $B=125$ kHz, then $\Delta_f=18.3$ kHz offset is equivalent to $\epsilon=149.91$ UI.

Under both timing and frequency offsets, \eqref{rn} is rewritten as
\begin{multline} \label{rn2}
r[n] = \sum_{l=0}^{L-1} h_{\rm RC}(l - \tau) p[n + N_{\rm CP}-l] \exp\left\{\frac{j 2 \pi n \epsilon}{M} \right\} + \eta[n], \\
n = 0, 1, \ldots, MN_{\rm PR}+N_{\rm CP}-1.
\end{multline}
Let $r_b[n]$ denote the received signal corresponding to the $b$th preamble symbol after removing the cyclic prefix. It can be expressed as
\begin{eqnarray} \label{rbn}
&r_b[n] &= \sum_{l=0}^{L-1} h_{\rm RC}(l - \tau)x_0^*[n-l] \exp\left\{j 2 \pi \frac{(n + Mb)\epsilon}{M} \right\} \nn \\
&       & + {\eta}[n + Mb] , \nn \\
&       &   n = 0, 1, \ldots, M-1, \; b= 0,1, \ldots, N_{\rm PR}-1
\end{eqnarray}

Performing de-chirping on $r_b[n]$ yields
\begin{eqnarray}
& u_b[n] & =\sum_{l=0}^{L-1} h_{\rm RC}(l-\tau) \exp\left\{-j \pi \left( \frac{l^2 + l}{M} \right) \right\} \nn \\
&        & \hspace{-1cm} \times \exp\left\{ \frac{j 2 \pi nl}{M}  \right\} \exp\left\{j 2 \pi \frac{(n + Mb)\epsilon}{M} \right\} + \hat{\eta}[n +
Mb] \nn \\
&        & = \sum_{l=0}^{L-1} h_{\rm RC}(l-\tau) \exp\left\{-j \pi \left( \frac{l^2 + l}{M} \right) \right\}  \nn \\
&        & \hspace{-0.3cm} \times \exp\left\{ \frac{j 2 \pi n(l+\epsilon)}{M} \right\} \exp\left\{j 2 \pi b\epsilon \right\} + \hat{\eta}[n + Mb].\nn\\
\end{eqnarray}
Next, $u_b[n]$ is processed by the DFT to produce
\begin{eqnarray} \label{vkb}
& v_b[k]& = \frac{1}{M}\sum_{l=0}^{L-1} h_{\rm RC}(l - \tau) \exp\left\{-j \pi \left(\frac{l^2}{M} + l \right) \right\}  \nn \\
& & \hspace{-1cm} \times \sum_{n=0}^{M-1} \exp\left\{\frac{-j 2 \pi n(k-\epsilon-l)}{M} \right\} \exp\left\{j 2 \pi b\epsilon \right\} +  \bar{\eta}_b[k]  \nn \\
\end{eqnarray}

For the sake of simplicity, define
\begin{equation} \label{hh}
\hat{h}_{\rm RC}(l-\tau) = h_{\rm RC}(l - \tau) \exp\left\{-j \pi \left(\frac{l^2}{M} + l \right)  \right\},
\end{equation}
which is a pulse whose magnitude is the same as the magnitude of the RC pulse. Furthermore, define
\begin{eqnarray}\label{psinc}
&\hspace{0cm}\Upsilon(k - \epsilon) &= \frac{1}{M}\sum_{n=0}^{M-1} \exp\left\{\frac{-j 2 \pi n(k-\epsilon)}{M} \right\} = \nn \\
& &\hspace*{-2.0cm}
\begin{cases}
 \frac{\sin(\pi (k-\epsilon))}{M \sin(\pi (k-\epsilon)/ M)} \exp\left\{-j
\pi \left( k-\epsilon - \frac{k-\epsilon}{M} \right) \right\}, & \text{if
} k \ne \epsilon, \\
 1 & \text{otherwise},
\end{cases}\nn\\
\end{eqnarray}
which is a periodic sinc function \cite{Nguyen16}, whose shape also resembles an RC pulse with infinitesimal roll-off factor. In fact, it can be easily seen that $\Upsilon(k - \epsilon)$ is an all-pass response that introduces a sample delay by an amount of $\epsilon$.

Then substituting \eqref{hh} and \eqref{psinc} into \eqref{vkb} yields
\begin{eqnarray} \label{vkb2}
&v_b[k] &= \sum_{l=0}^{L-1} \hat{h}_{\rm RC}(l-\tau) \Upsilon(k-\epsilon -l) \exp\left\{j 2 \pi b\epsilon \right\} +  \bar{\eta}_b[k] \nn \\
& & \approx \hat{h}_{\rm RC}(l-\tau-\epsilon) \exp\left\{j 2 \pi b\epsilon \right\} +  \bar{\eta}_b[k].
\end{eqnarray}
An important and useful observation of \eqref{vkb2} is that, in the absence of noise, the peak magnitude of $v_b[k]$ is shifted by an amount equivalent to $\tau+\epsilon$ when both timing and frequency offsets exist.

One simple way to separate these two offsets is to use a preamble sequence contains both \emph{down-chirp} and \emph{up-chirp} symbols. Specifically, performing the same signal processing operations as in \eqref{rn2}-\eqref{vkb2} on up-chirp preamble symbols yields a signal whose magnitude follows the magnitude of an RC pulse, but shifted by an amount $\epsilon-\tau$. Therefore, by combining peak detection results from both down-chirp
and up-chirp preamble symbols, we can find timing and frequency offsets separately. This is described in more detail in the following.

Let $N_{\rm D}$ and $N_{\rm U}$, respectively, denote the numbers of down-chirp and up-chirp symbols. First, the DFT outputs of all the down-chirp
preamble symbols are summed together in terms of power to obtain:
\begin{equation}
V_{\rm D}[k] = \sum_{b=0}^{N_{\rm D}-1} |v_b[k]|^2,
\end{equation}
from which the timing delay $\tau_{\rm D}$ is calculated based on the location of the peak sample. Similarly, the DFT outputs of all the up-chirp preamble symbols are summed together to give
\begin{equation}
V_{\rm U}[k] = \sum_{b=N_{\rm D}}^{N_{\rm D}+N_{\rm U}-1} |v_b[k]|^2,
\end{equation}
from which the timing delay $\tau_{\rm U}$ is also calculated. It follows
that the coarse estimates of timing and frequency offsets can be obtained
as
\begin{equation}\label{tau-coarse}
\tau_{\rm coarse} = 0.5 ( \tau_{\rm D} - \tau_{\rm U}) \quad (\text{UI}),
\end{equation}
and
\begin{equation}\label{epsilon-coarse}
\epsilon_{\rm coarse} = 0.5 ( \tau_{\rm D} + \tau_{\rm U}) \quad (\text{UI}),
\end{equation}

\begin{figure*}[htb!]
\centering
\includegraphics[scale=0.80]{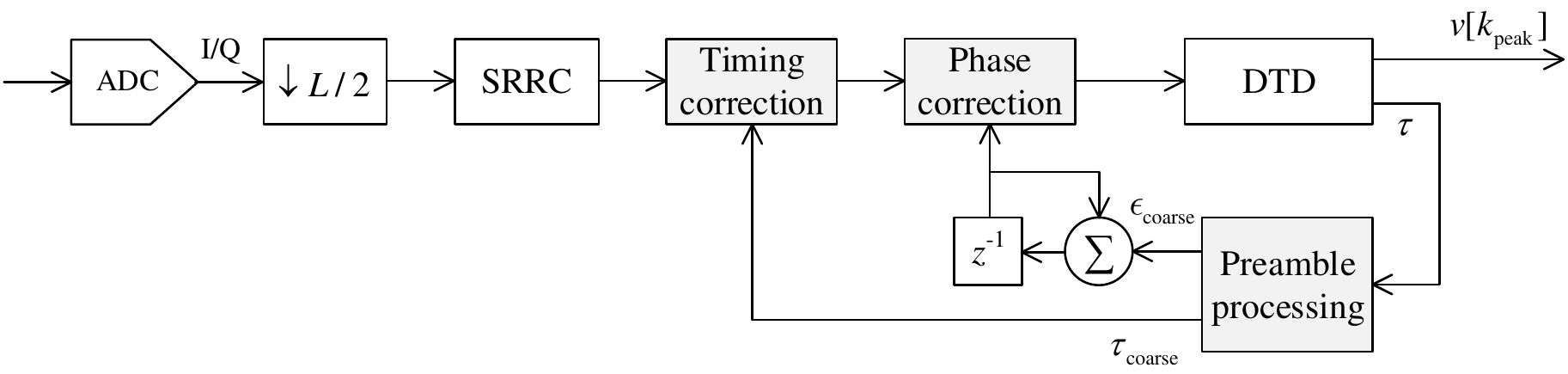}
\caption{Block diagram of the proposed non-coherent receiver.}
\label{fig-noncoh-receiver}
\end{figure*}

With the coarse estimates of the timing and frequency offsets, a complete block diagram of the proposed non-coherent receiver design
is shown in Fig. \ref{fig-noncoh-receiver}. Specifically, the DTD block contains the demodulator and timing detector modules as detailed in Fig. \ref{fig-detector}, which produces raw timing delay $\tau$ for each preamble symbol. From the raw timing delay, the preamble processing block produces coarse timing and frequency errors, i.e., \eqref{tau-coarse} and \eqref{epsilon-coarse}. These coarse timing and frequency errors are corrected before making non-coherent detection of the data symbols. Note that the detected frequency error is accumulated from sample to sample to create a continuous linear phase ramp, which is used for frequency correction.

While the main ideas and algorithms presented in this paper were originally described in \cite{CSS-TxRx}, a similar idea of using both up chirps and down chirps in the preamble for frame synchronization of LoRa signals is recently presented in \cite{Bernier20}. It should be pointed out, however that the timing, frequency and phase detection techniques developed in this paper are different from what presented in \cite{Bernier20}. More importantly, while the work in \cite{Bernier20} is mainly concerned with preamble detection performance and frame synchronization failure probability, the present paper addresses not only coarse timing and frequency synchronization, but also fine timing, frequency and phase synchronization based on data-directed symbols. This fine synchronization is presented in the next subsection, which plays a key role in achieving the excellent BER performance of the recently proposed PSK-CSS scheme.

\subsection{Fine Timing, Frequency and Phase Synchronization for the Proposed Coherent Receiver}\label{sec-fine}

While the solutions for \emph{coarse} timing and frequency synchronization in \eqref{tau-coarse} and \eqref{epsilon-coarse} can be used in the proposed non-coherent receiver, further refinement can be achieved via \emph{fine} timing and frequency synchronization that makes use of data-directed symbols. More importantly, accurate phase detection of data symbols is also achieved, which facilitates the practical implementation of the higher-rate PSK-CSS scheme proposed in \cite{Nguyen19,Bom19}.

\begin{figure*}[htb!]
\centering
\includegraphics[scale=0.80]{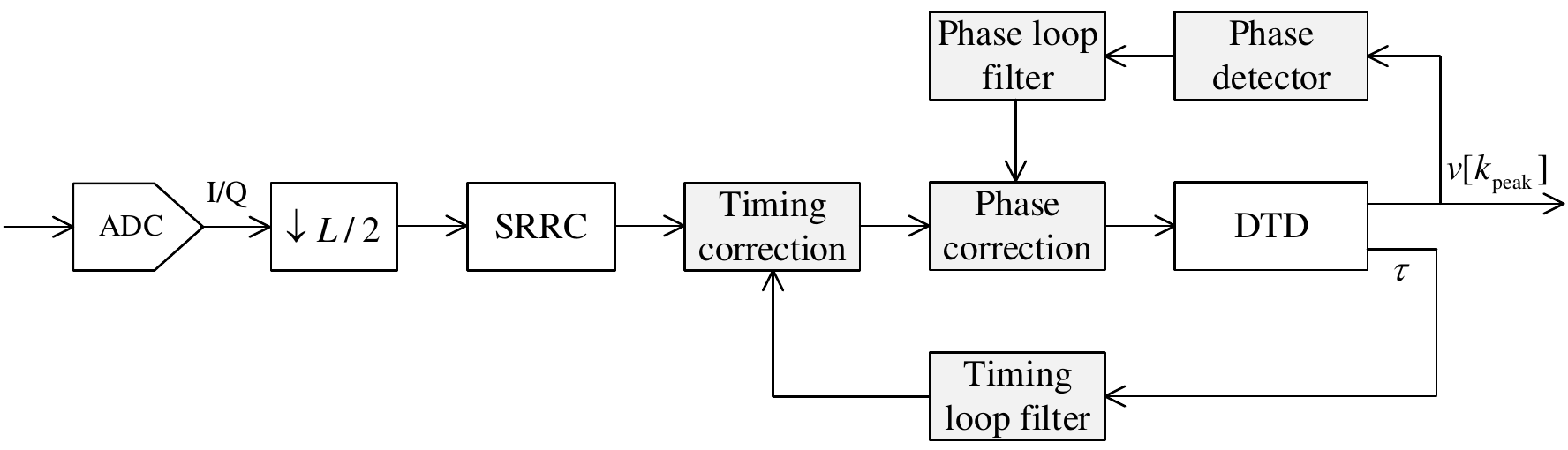}
\caption{Block diagram of the proposed coherent receiver.}
\label{fig-receiver}
\end{figure*}

A complete block diagram of the proposed coherent receiver design is shown in Fig. \ref{fig-receiver}. As before, the DTD block contains the demodulator and timing detector modules as detailed in Fig. \ref{fig-detector}. The receiver has a \emph{timing loop} to track the timing difference between the transmitter and the receiver using all the transmitted symbols, including preamble and data, in a burst. The detected timing error from the DTD module is sent to a \emph{loop filter} that drives the \emph{timing correction module}. The loop filter is designed as a first-order phase-locked loop (PLL) with a transfer function of
\begin{equation} \label{loop1}
H^{(1)}(z) = \frac{k_g z^{-1}}{1-z^{-1}}
\end{equation}
where $k_g$ is the proportional loop gain. The timing correction module can be implemented using the well-known Farrow filter structure, or any timing interpolation methods in the literature \cite{Ves96}. The loop gain is an important factor that influences the receiver's performance. In particular, the gain provides a trade-off between convergence speed versus timing correction accuracy. A higher gain makes the loop converges faster,
whereas a lower gain is more effective at reducing the phase detector's noise.

\begin{figure}[htb!]
\centering
\includegraphics[scale=0.8]{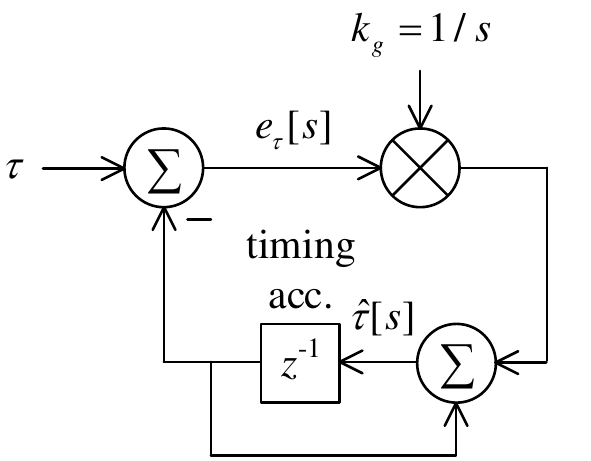}
\caption{First-order phase locked loop for tracking timing.}
\label{fig-1st-order}
\end{figure}

Fig. \ref{fig-1st-order} shows a simplified model of the timing loop filter, in which the timing offset between the transmitter and the receiver
is denoted as $\tau$. The PLL runs at the CSS symbol rate, which is $1/M$
of the CSS sample rate since there are $M$ samples in one CSS symbol. As shown in the figure, the first-order loop contains a single accumulator whose value indicates the best estimate of the timing, hence the name \emph{timing accumulator}. Let $s=1,2,3\ldots$ denote indices of CSS symbols in a CSS burst. At each iteration, the estimated time $\hat{\tau}[s]$ is applied to the next symbol using a timing correction circuit as shown in Fig. \ref{fig-receiver}. Since timing detection is affected by noise and tracking performance of the loop itself, the detected timing error can be modelled as
\begin{equation}
e_\tau[s] = \tau - \hat{\tau}[s-1] + w_\tau[s],
\end{equation}
where $w_\tau[s]$ is timing detection noise. To accelerate loop convergence, the timing accumulator is seeded with the coarse timing detected by processing the
preamble, i.e.,
\begin{equation}
\hat{\tau}[0] = \tau_{\rm coarse}.
\end{equation}
The loop employs a gain factor $k_g$ to filter out the detection noise. It can be shown that the best
and unbias estimate can be obtained by using the gain factor
\begin{equation}
k_g = \frac{1}{s}, \quad s=1,2,3,\ldots
\end{equation}

While a formal proof of the above gain can be carried out by applying the
Kalman filter formula on the PLL circuit, it can be intuitively demonstrated by the following operations of the circuit in Fig. \ref{fig-1st-order}:
\begin{itemize}
	\item At the beginning, $\hat{\tau}[0] = 0$ and thus $e_\tau[1] = \tau + w_\tau[1]$.
	\item The accumulator is loaded with the new value: $\hat{\tau}[1] = \tau + w_\tau[1]$.
	\item The next symbol comes in, and a new timing error is detected: $e_\tau[2] = \tau + w_\tau[2] - (\tau + w_\tau[1]) = w_\tau[2] - w_\tau[1]$.
	\item The accumulator's new value is: $\hat{\tau}[2] = \tau + w_\tau[1] + \frac{w_\tau[2] - w_\tau[1]}{2} = \tau + \frac{w_\tau[2] + w_\tau[1]}{2}$.
	\item The next timing error is: $e_\tau[3] = \tau + w_\tau[3] - \tau -
\frac{w_\tau[2] + w_\tau[1]}{2} = w_\tau[3] - \frac{w_\tau[2] + w_\tau[1]}{2} $
	\item The accumulator's next value is: $\hat{\tau}[3] = \tau + \frac{w_\tau[2] + w_\tau[1]}{2} + \frac{w_\tau[3] - \frac{w_\tau[2] + w_\tau[1]}{2}}{3} =  \tau + \frac{w_\tau[3] + w_\tau[2] + w_\tau[1]}{3}$.
\end{itemize}
Therefore, it can be established iteratively that
\begin{equation} \label{hattaus}
\hat{\tau}[s] = \tau  + \frac{1}{s}\sum_{j=1}^s w_\tau[j],
\end{equation}
showing that the timing estimate is \emph{unbiased}, and improved at every iteration. This leads to large performance improvement as compared to a
conventional PLL having a fixed loop gain.

Finally, the receiver also contains a frequency-and-phase tracking loop that tracks the differences in frequency and phase between the transmitter
and the receiver using all the transmitted symbols in a CSS packet. The tracking loop contains a \emph{phase detector}, a \emph{phase loop filter}
and a \emph{phase correction circuit}. The phase error is extracted from the angle of $v_s[k_{\rm peak}]$ as
\begin{equation} \label{phi}
e_\phi[s] = {\tt smod} \left( \frac{\angle{ v_s[k_{\rm peak}] }}{ 2 \pi}, 1/Q \right) \quad ({\rm UI}),
\end{equation}
where $Q$ denotes the order of the PSK modulation employed for the initial phase of each chirp (e.g. $Q=4$ for QPSK), and {\tt smod} denotes the
signed modulo function, defined as
\begin{equation}
{\tt smod}(a, b) = a - {\tt round}(a/b) \times b.
\end{equation}

The output of the phase detector is sent to a phase loop filter to drive the phase correction circuit. In order to track both frequency and phase,
the loop filter is designed as a second-order PLL with a transfer function of
\begin{equation} \label{loop2}
H^{(2)}(z) = \left(k_p + \frac{k_i}{1-z^{-1}} \right) \frac{z^{-1}}{1-z^{-1}}
\end{equation}
where $k_p$ and $k_i$ are proportional gain and integral gain, respectively.

\begin{figure}[htb!]
\centering
\includegraphics[scale=0.80]{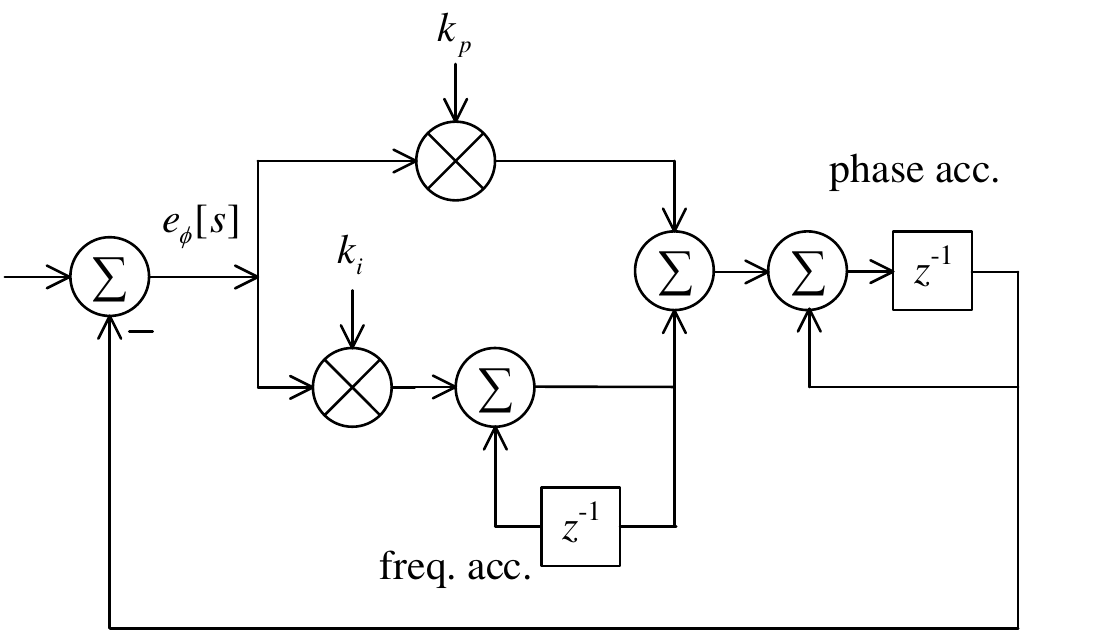}
\caption{Second-order phase locked loop for tracking frequency and phase.}
\label{fig-2nd-order}
\end{figure}

Fig. \ref{fig-2nd-order} shows a simplified model of the phase loop system using a second-order PLL. Unlike the first-order loop, the second-order
loop contains two accumulators, namely frequency accumulator and phase accumulator. After each iteration, the values inside the accumulators represent the best estimates of frequency and phase at that particular moment.
To accelerate loop convergence, the frequency accumulator is initially seeded with the coarse frequency estimate, i.e., $\epsilon_{\rm coarse}$, and the phase accumulator is seeded with the phase extracted from the last
preamble symbol.

Similar to the timing loop, the phase loop can be optimized by representing the second-order PLL as a Kalman filter \cite{Patap99}. Specifically, define a dynamic system model with a state transition function of
\begin{equation}
\mathbf{X}_{s+1} = \mathbf{F} \mathbf{X}_s + \mathbf{W}_s,
\end{equation}
and an observation function of
\begin{equation}
\phi[s] = \mathbf{H} \mathbf{X}_s + w_\phi[s],
\end{equation}
where
\begin{itemize}
	\item $\mathbf{X}_s$ is the hidden state, estimated at the $s$th iteration.
	\item $\mathbf{F} = \begin{bmatrix} 1 & 1 \\ 0 & 1 \end{bmatrix}$ is the state transition matrix.
	\item $\mathbf{W}_s$ is the random process noise.
	\item $\mathbf{H} = \begin{bmatrix} 1 & 0 \end{bmatrix}$ is the observation matrix.
	\item $w_\phi[s]$ is the observation noise, i.e., the phase detector's noise.
\end{itemize}

From the above model, the Kalman gains for the second-order PLL can be iteratively calculated as
\begin{equation}
\begin{bmatrix} k_p[s] \\ k_i[s] \end{bmatrix} = \mathbf{K}_s = \frac{\mathbf{P}_s \mathbf{H}^T}{\mathbf{H} \mathbf{P}_s \mathbf{H}^T + \sigma^2_\phi }.
\end{equation}
It then follows that a \emph{prediction} of the measurement can be calculated as
\begin{equation}
\hat{\phi}[s] = \mathbf{H} \hat{\mathbf{X}}_s,
\end{equation}
and an estimation of the new state is
\begin{equation}
\hat{\mathbf{X}}_{s+1} = \hat{\mathbf{X}}_s + \mathbf{K}_s (\phi[s] - \hat{\phi}[s]) = \hat{\mathbf{X}}_s + \mathbf{K}_s e_{\phi}[s].
\end{equation}
Furthermore, the predicted state covariance matrix is calculated as
\begin{equation}
\mathbf{P}_{s+1} = \mathbf{F}( \mathbf{I} - \mathbf{K}_s \mathbf{H})  \mathbf{P}_s \mathbf{F}^T + \mathbf{Q}.
\end{equation}
In the above expression, $\mathbf{Q}$ is the process noise covariance matrix, given as
\begin{equation}
\mathbf{Q} = \begin{bmatrix} \sigma^2_{\eta}  & 0\\ 0 & \sigma^2_{\zeta} \end{bmatrix},
\end{equation}
where $\sigma^2_{\eta} $ and $\sigma^2_{\zeta}$ are variances of the phase and frequency disturbances of the local oscillator. These variances are
typically characterized by the phase noise analysis during the circuit calibration process.

It is important to seed values of the predicted covariance matrix at a proper condition prior to the first iteration, which has the form
\begin{equation}
\mathbf{P}_0 = \begin{bmatrix} \sigma^2_{\phi}  & 0\\ 0 & \sigma^2_{f} \end{bmatrix}.
\end{equation}
To this end, the phase variance is approximated from the complex Gaussian
noise affecting $v_s[k_{\rm peak}]$ as \cite{Berscheid11}:
\begin{equation}
\sigma^2_{\phi} = 0.5 \frac{10^{-{\rm PSNR}/10}}{ (2 \pi)^2}.
\end{equation}
On the other hand, the frequency variance can be shown to be inversely proportional to the number of preamble symbols employed, i.e.,
\begin{equation}
\sigma^2_f = \frac{\sigma^2_{\phi}}{N_{\rm D}+N_{\rm U}}.
\end{equation}

\section{Simulation Results}\label{sec-res}
Performance of the proposed non-coherent and coherent receivers is investigated with Monte Carlo simulations in this section. First, convergence performance of the timing loop and the phase loop is evaluated, followed by the BER performance of the complete system under realistic channel conditions. In all simulations, the signal bandwidth is selected as 125 kHz, the channel is assumed to have a constant gain with an unknown delay and additive white Gaussian noise. Each of CSS bursts has 256 data symbols, preceded by a preamble having $N_{\rm D}=8$ down chirps and $N_{\rm U}=8$ up chirps. Since the detection of integer timing and frequency errors (i.e., the amount of timing or frequency error at an integer multiples of
1 UI) is budgeted as part of burst (frame) detection \cite{Bernier20}, the channel model in this section is assumed to introduce fractional errors
only, where both $\tau$ and $\epsilon$ are uniformly distributed between $-0.5$ and $0.5$ UI. The digital models of both transmitter and receiver operates with the up-sampling factor $L=2$. The SRRC filters are selected to have the roll-off factor of 0.25 and designed as truncated finite impulse response (FIR) filters with 33 coefficients, i.e., the filter order is 16. At the receiver side, the timing correction block is designed as
a the fifth-order Farrow interpolator that runs at an over-sampling factor of 2.

\subsection{Convergence of the Timing Loop}

\begin{figure}[htb!]
\centering
\includegraphics[width=0.45\textwidth]{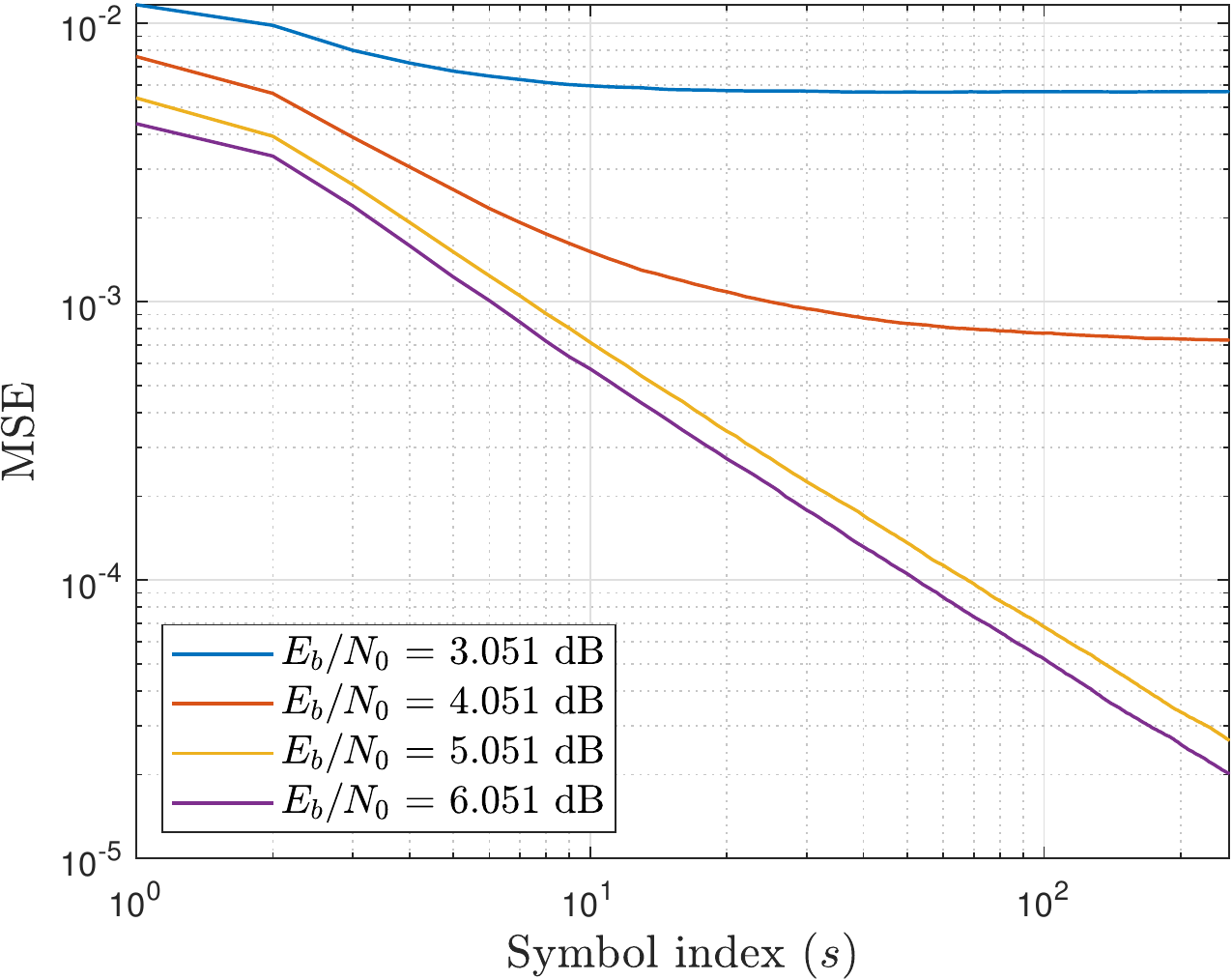}
\caption{Convergence performance of the timing loop.}
\label{fig-timing-mse}
\end{figure}

Fig. \ref{fig-timing-mse} plots simulation results illustrating convergence of the timing loop with ${\rm SF}=8$ and four different SNR values, namely of $-12$, $-11$, $-10$ and $-9$ dB. These SNR values correspond to
$E_b/N_0=3.051$, $4.051$, $5.051$, and $6.051$ dB. The detected timing,
which is the output of the timing loop filter, is recorded after each symbol and compared with the true channel delay to determine the mean square
error (MSE) of the timing detection. It can be seen that the accuracy of the proposed timing detector gradually gets better after each detected symbol, thanks to the timing loop filter. When the SNR is above a certain threshold, e.g., $E_b/N_0 > 5$ dB, the MSE curves decline at a rate of $-10$ dB per decade, which matches the theoretical prediction in \eqref{hattaus}. On the other hand, when the SNR is below the threshold (i.e., the cases $E_b/N_0 = 3.051$ and $E_b/N_0=4.051$), the MSE curves approach error floors. The reason for this is that, at low SNR, the peak detection
$k_{\rm peak}$ in the first branch of the DTD is not so reliable, causing
large errors in timing detection in the second branch. However, the case that $E_b/N_0 < 5 $ dB is not common in practical settings of LoRa systems (see, e.g., \cite{Elsha18,Nguyen19,Bernier20}) since it leads to poor BER for $SF=8$. Therefore the MSE floors exhibited in \ref{fig-timing-mse} shall not be a concern.

\subsection{Convergence of the Phase Loop}

\begin{figure}[htb!]
\centering
\includegraphics[width=0.45\textwidth]{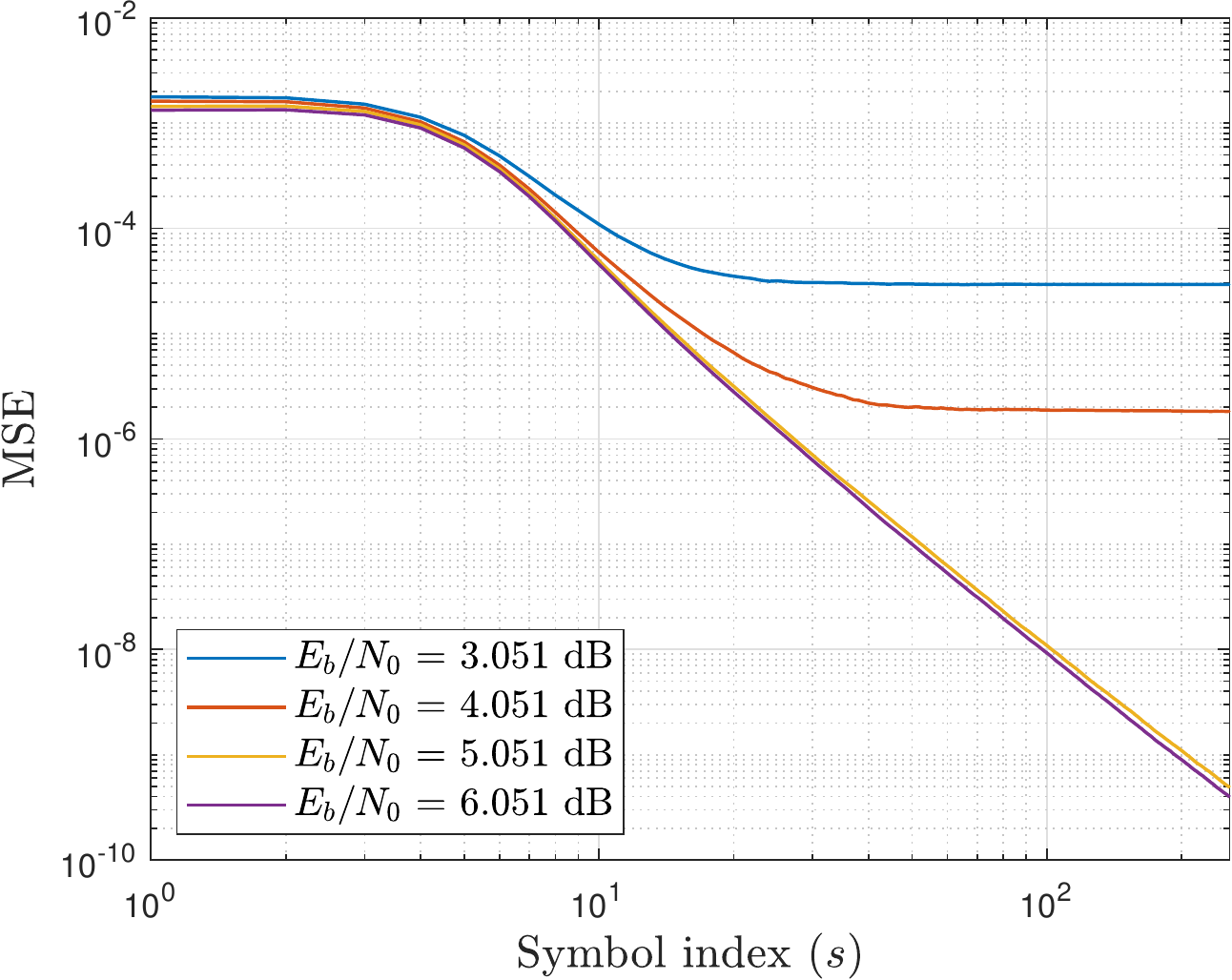}
\caption{MSE in tracking frequency.}
\label{fig-freq-mse}
\end{figure}

\begin{figure}[htb!]
\centering
\includegraphics[width=0.45\textwidth]{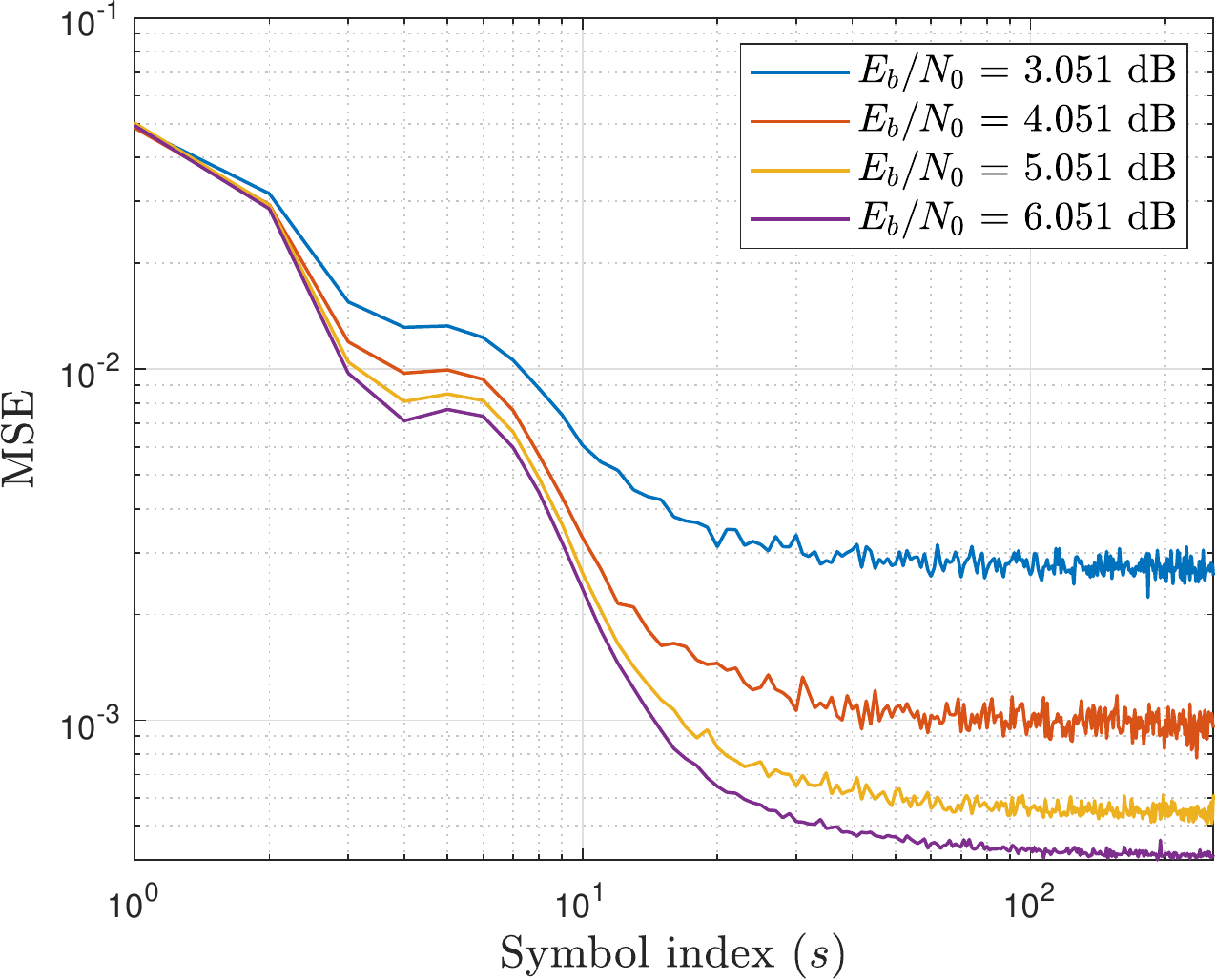}
\caption{MSE in tracking phase.}
\label{fig-phase-mse}
\end{figure}

Figs. \ref{fig-freq-mse} and \ref{fig-phase-mse} show typical frequency and phase convergence curves, respectively, of the phase tracking loop for
the same CSS system with ${\rm SF}=8$. The frequency error is obtained by comparing the value of the frequency accumulator inside the phase loop
filter against the actual frequency error introduced by the channel, whereas the phase error is obtained at the output of the phase detector. It can be seen that the MSE of the phase estimate reduces at the rate of about $-10$ dB per decade before it reaches an error floor after about 16 symbols. The error floor is equivalent to $\sigma^2_{\phi}$, which agrees with the theoretical analysis. Since frequency, by definition, is the derivative of phase, the convergence rate of the frequency MSE is $-20$ dB per
decade, which can be seen from Fig. \ref{fig-freq-mse}. Again, the frequency MSE floors observed for $E_b/N_0 = 3.051$ and $4.051$ dB are due to
erroneous peak detection at low SNR, a phenomenon similar to what observed with the timing MSE plot in Fig. \ref{fig-timing-mse}.

\subsection{Performance of the Proposed Non-Coherent Receiver}

\begin{figure}[htb!]
\centering
\includegraphics[width=0.45\textwidth]{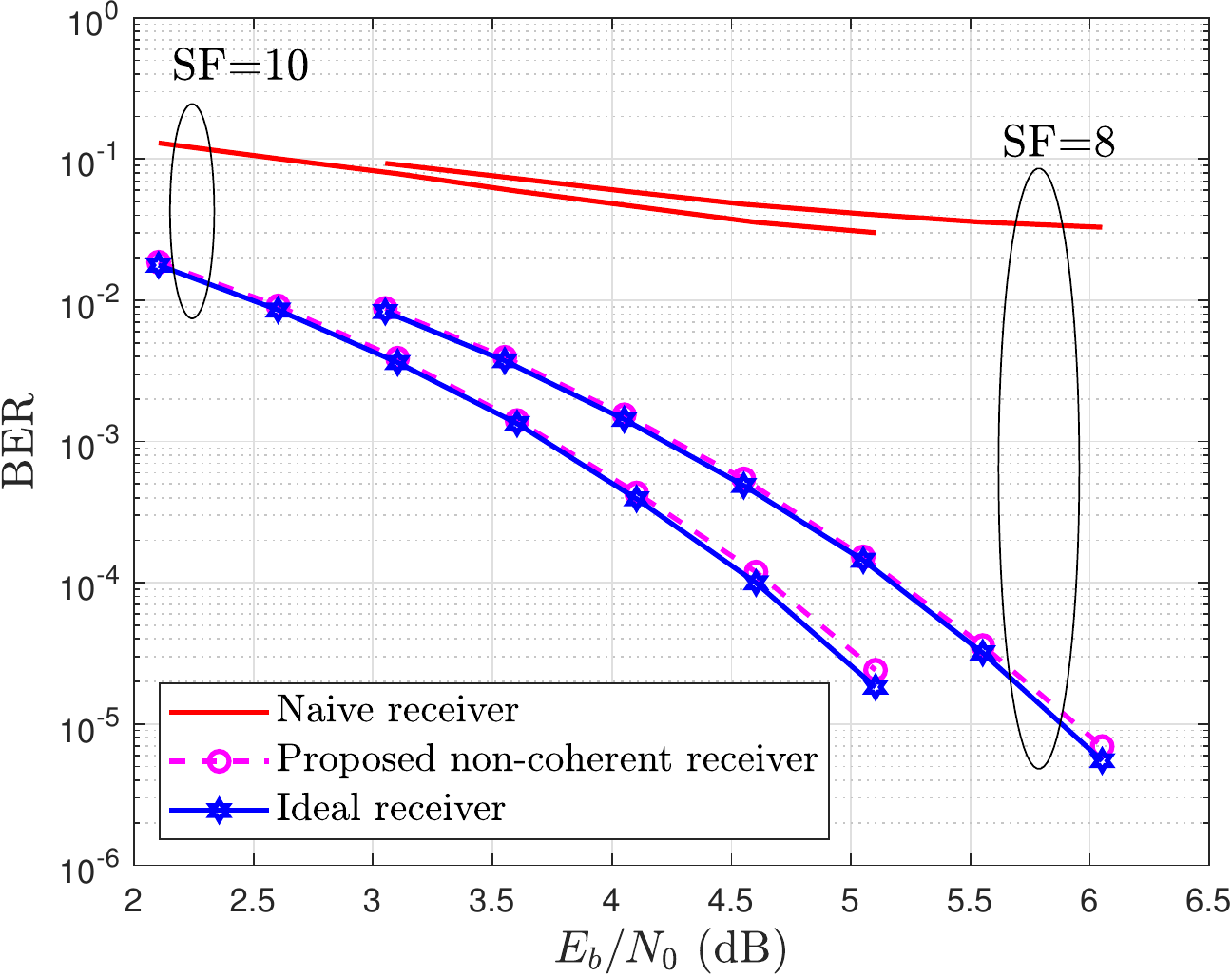}
\caption{BER performance of the CSS system with the proposed non-coherent
receiver.}
\label{fig-noncoh-det}
\end{figure}

The BER performance of the proposed non-coherent receiver in Fig. \ref{fig-noncoh-receiver} is illustrated in Fig. \ref{fig-noncoh-det} for two different spreading factors ${\rm SF}=8$ and ${\rm SF}=10$. To show the
importance of timing and frequency synchronization, performance of the \emph{naive} receiver is also shown in this figure. The naive receiver is a
non-coherent receiver that does not include neither the timing nor frequency correction module, and is evaluated under the presence of random timing and frequency errors. As expected, performance of such a naive non-coherent receiver is very poor (unacceptable) without timing and frequency synchronization. On the contrary, as can be seen from Fig. \ref{fig-noncoh-receiver}, the proposed non-coherent receiver implementing only coarse timing and frequency detections enjoys the BER performance very close to the performance of an ideal non-coherent receiver.

\subsection{Performance of the Proposed Coherent Receiver}

Having demonstrated the convergence performance of timing, frequency and phase synchronization, this section presents the BER performance of the proposed coherent receiver for two different spreading factors ${\rm SF}=8$ and ${\rm SF}=10$.

\begin{figure}[htb!]
\centering
\includegraphics[width=0.45\textwidth]{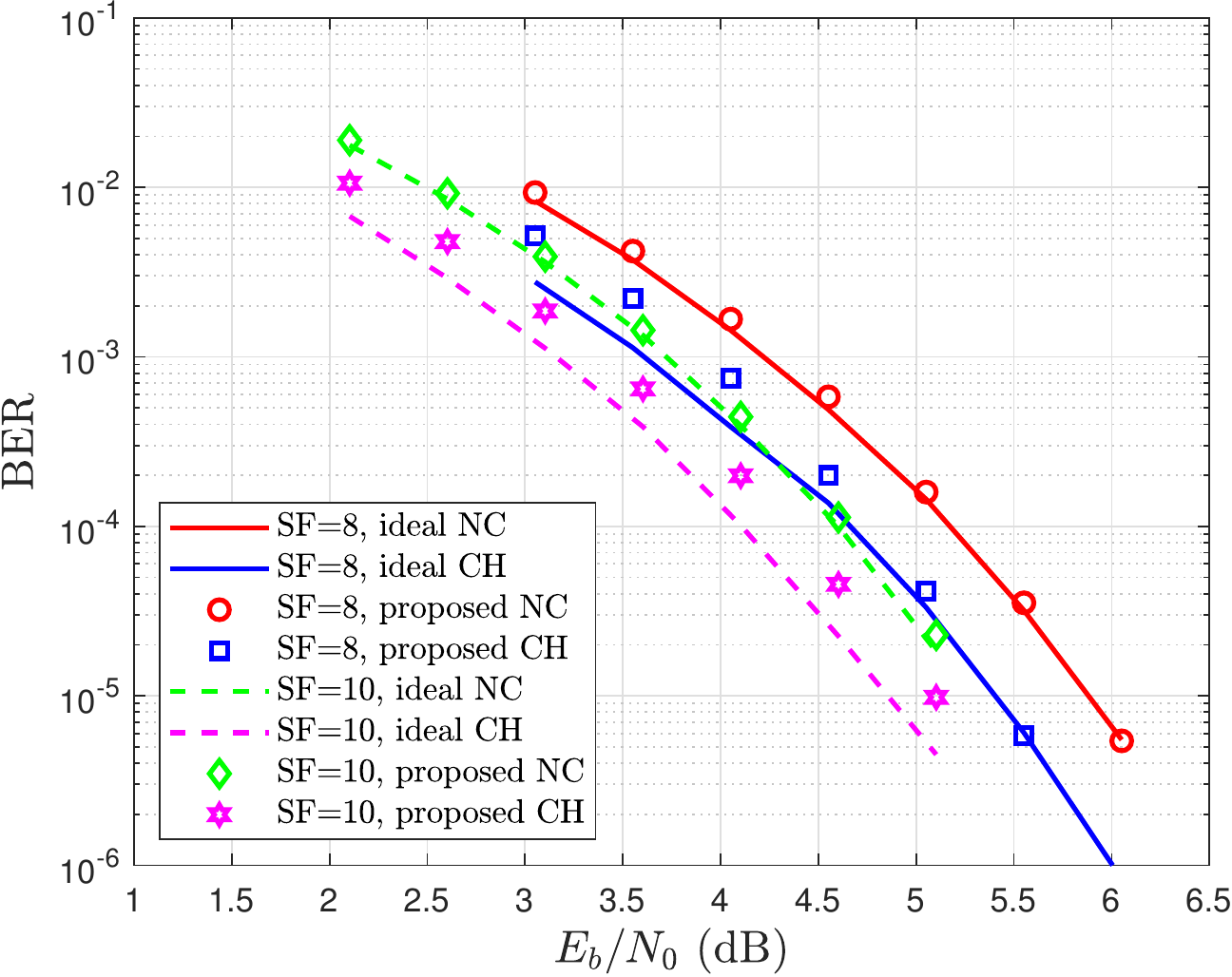}
\caption{BER performance of the CSS system with the proposed coherent receiver.}
\label{fig-coh-det}
\end{figure}

First, BER performance comparison between coherent (CH) and non-coherent (NC) detection methods is shown in Fig. \ref{fig-coh-det} in the presence
of fractional timing and frequency errors. The system parameters used in the simulation are ${\rm SF}=8$, $N_{\rm D}=N_{\rm U}=8$. Each transmission burst contains 256 CSS symbols. For each burst, both the timing and frequency errors are uniformly distributed between $-0.5$ and $0.5$ UI. Because coherent detection relies on good phase tracking performance, in practice, it can be turned on after the phase loop filter is fully converged. Observation of the phase MSE plot in Fig. \ref{fig-phase-mse} suggests that phase convergence is achieved after 16 symbols. For that reason, the first 16 symbols are detected using the non-coherent method while the remaining symbols of the data burst (i.e., 240 CSS symbols) are detected using the proposed coherent method. To see how well the proposed receiver performs, the BER performance of the \emph{ideal} non-coherent receiver and \emph{ideal} coherent receiver is also shown in the figure. The ideal coherent receiver performs ideal timing, frequency and phase compensation assuming perfect knowledge of the errors introduced by the channel.

It can be seen from Fig. \ref{fig-coh-det} that the ideal coherent receiver provides about 0.5 dB gain compared to the ideal non-coherent receiver
for both cases of ${\rm SF}=8$ and ${\rm SF}=10$. With ${\rm SF}=8$
and $E_b/N_0 > 5$dB, the proposed receiver yields almost the same performance as that of the ideal counterpart, in both cases of non-coherent an coherent detection. The observation that the proposed receiver with non-coherent detection performs identical to the ideal con-coherent receiver indicates that both the timing and frequency loops perform well enough to a
point that fractional timing and frequency error are no longer a performance limiting factor. For ${\rm SF}=10$, the proposed coherent receiver performs about 0.25 dB worst than the ideal coherent receiver. While such
a performance gap is small enough to justify the practical use of the proposed coherent receiver, it also suggests there is a room for improvement
when it comes to phase synchronization at higher SF values.

\begin{figure}[htb!]
\centering
\includegraphics[width=0.45\textwidth]{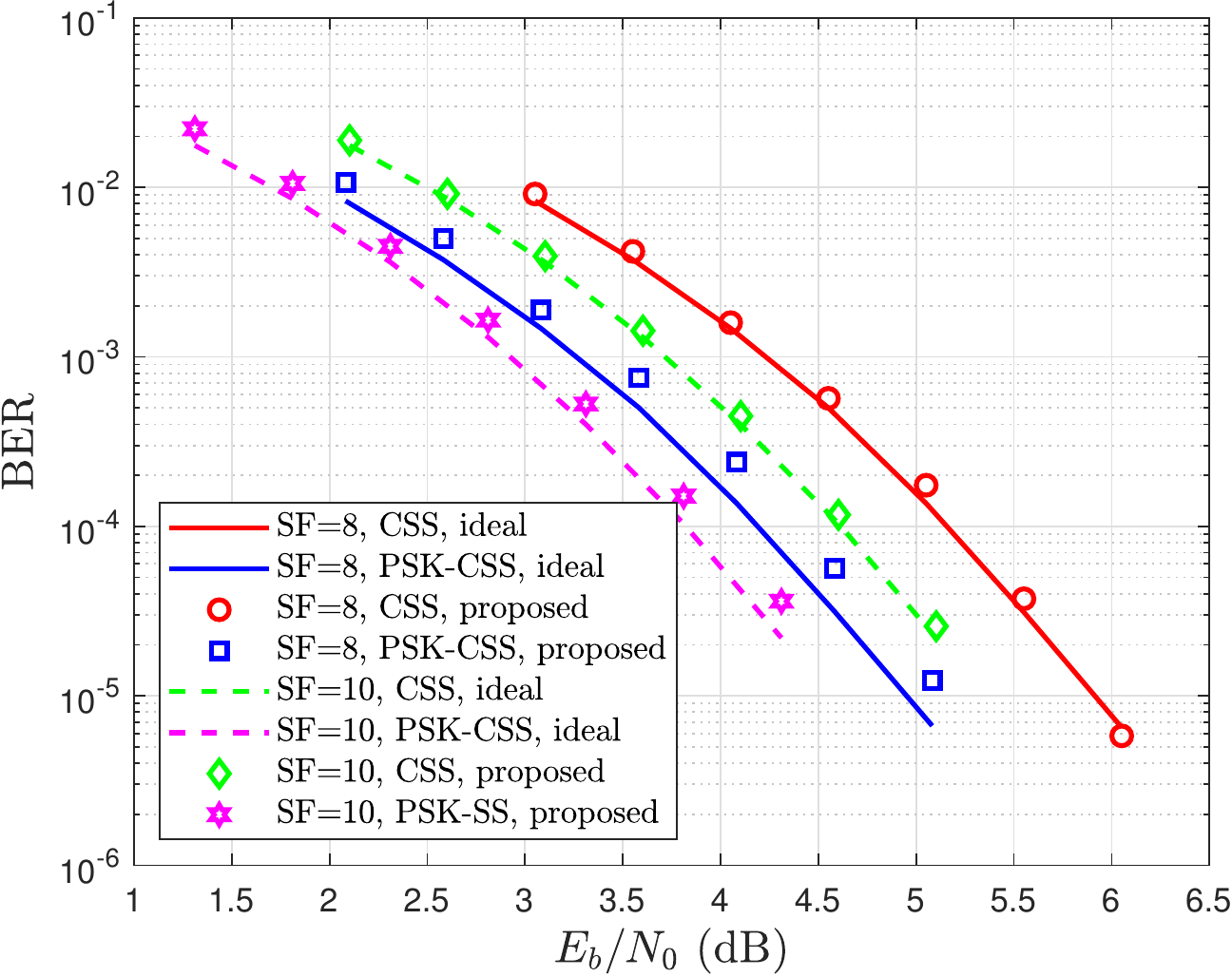}
\caption{BER performance of the PSK-CSS scheme with the proposed coherent
receiver.}
\label{fig-pscss-det}
\end{figure}

Finally, Fig. \ref{fig-pscss-det} plots the BER performance of the PSK-CSS system proposed in \cite{Nguyen19} under a realistic channel condition with fractional timing and frequency errors. Among the 256 CSS symbols on
each burst, the first 16 symbols are not phase modulated (i.e., $Q=1$) considering the fact that phase synchronization is not fully converged in
the first 16 symbols (see Fig. \ref{fig-phase-mse}). The remaining 240 data symbols carry extra information bits embedded in the initial phases of
CSS symbols. Using QPSK modulation (i.e., $Q=$ 4) allows to send two additional bits per each of these 240 CSS symbols. With ${\rm SF}=8$, this represents $\frac{(8\times 16+ 10\times 240)}{8\times 256}=123.44\%$ data rate improvement, whereas it is $118.75\%$ for ${\rm SF}=10$. As can be seen, the ideal receiver with perfect synchronization provides 1.0 and 0.8 dB gains over the ideal non-coherent receiver of the conventional
CSS system for ${\rm SF}=8$ and ${\rm SF}=10$, respectively. More importantly, the BER performance of the PSK-CSS scheme achieved with the proposed practical coherent receiver has only 0.25 dB gaps as compared to the ideal co-coherent receiver for both considered SF values.

\section{Conclusions}\label{sec-con}

With the objective of designing non-coherent and coherent receivers that can work under realistic channel conditions of having timing and frequency offsets, this paper has developed high-performance synchronization methods for recovering timing, frequency and phase of CSS signals. A novel method was proposed to perform coarse timing and frequency synchronization that makes use of the preamble consisting of both down chirps and up chirps and the shape of the pulse shaping/matched filters. Furthermore, to enable fine synchronization, timing and phase loop filters were designed as
simple first-order and second-order PLL circuits with dynamic gain control.

The obtained results show that, in practical scenario, timing and frequency synchronization is essential to non-coherent detection of CSS signals.
The proposed non-coherent receiver, with only coarse timing and frequency
detections, performs almost identical to an ideal non-coherent receiver. By further employing the fine timing, frequency and phase synchronization, the proposed coherent receiver yields about 0.5 dB gain as compared to the proposed non-coherent receiver. Finally, impressive BER performance obtained with the proposed coherent receiver is demonstrated for the higher-rate PSK-CSS system, which provides about 0.75 dB as compared to non-coherent detection of conventional CSS signals, and at the same time delivering $123.44\%$ and $118.75\%$ data rate improvements for ${\rm SF}=8$ and ${\rm SF}=10$, respectively.

\section*{Acknowledgement}

This work is supported by the NSERC/Cisco Industrial Research Chair in Low-Power Wireless Access in Sensor Networks.

\bibliographystyle{ieeetr}

\balance
\end{document}